\documentclass[twocolumn,prb,showpacs]{revtex4}

\usepackage{amsmath,amssymb}
\usepackage{graphicx}
\usepackage{dcolumn}
\usepackage{bm}

\begin{document}



\title{Shot-noise and conductance measurements of transparent superconductor~/~two-dimensional electron gas junctions}

\author{B.-R.\ Choi}
\author{A.\ E.\ Hansen}
\author{T.\ Kontos}
\author{C.\ Hoffmann}
\author{S.\ Oberholzer}
\author{W.\ Belzig}
\author{C.\ Sch{\"o}nenberger}
\email{Christian.Schoenenberger@unibas.ch}
\homepage{www.unibas.ch/phys-meso}
\affiliation{Institute of Physics, University of Basel, Klingelbergstr. 82, CH-4056 Basel, Switzerland}

\author{T.\ Akazaki}
\author{H.\ Takayanagi}
\affiliation{NTT Basic Research Laboratories, 3-1 Morinosato-Wakamiya, Atsugi-shi, Kanagawa 243-0198, Japan}

\date{\today}

\begin{abstract}
We have measured the conductance and shot-noise of superconductor-normal metal
(S-N) junctions between a Niobium (Nb) film and a
2-dimensional electron gas (2DEG), formed in an InAs-based semiconductor he\-te\-ro\-struc\-ture.
Adjacent to the junction, the 2DEG is shaped into a sub-micrometer beam-splitter.
The current shot-noise measured through one arm of the beam-splitter is found to be enhanced due to
Andreev reflection. Both noise and conductance measurements indicate that the Nb-2DEG
interface is of high quality with a transparency approaching \mbox{$\approx 60-70$\,\%}. The
present device can be seen as a quasi-ballistic S-N beam-splitter junction.
\end{abstract}

\pacs{
74.78.Na, 
73.23.-b, 
73.63.Rt, 
74.45.+c, 
73.50.Td, 
72.70.+m  
}

\keywords{shot-noise,S-N junction,nanocontact,proximity effect}

\maketitle

\section{\label{int}Introduction}

Shot-noise measurements provide a powerful tool to study
charge transport in mesoscopic systems.\cite{Beenakker03} Whereas resistance
measurements yield information on the average
probability for the transmission of electrons from source to drain,
shot-noise provides additional information on the electron transfer process,
which can not be obtained from resistance measurements.
For instance, the charge of quasi-particles can be extracted from
shot-noise measurements, an experiment that was applied to the
fractional quantum Hall regime.\cite{Saminadayar97,dePicciotto97,Reznikov99}
Shot-noise also provides information on the statistics of the electron transfer.
In general, the fermionic nature of the particles lead to a suppression
of the shot-noise from its classical value $S_I=2e|I|$, corresponding to Poissonian statistics
($S_I$ is the power-spectral density of current fluctuations in units of \mbox{A$^2$s}).
Suppression can also be induced
by Coulomb interaction, which was observed in the single-electron tunneling
regime.\cite{Birk95} That shot-noise can be fully suppressed in an open channel
was confirmed in quantum-point contacts.\cite{Kumar96,Reznikov95}
In a general conductor, the suppression is not full, but depends on the actual
distribution of transmission eigenvalues.\cite{Khlus87,Lesovik89,Buettiker90}
For example, shot-noise is suppressed to $1/3$ in
a disorder wire~\cite{Beenakker92a,Nagaev92,Liefrink94,Steinbach96,Schoelkopf97,Henny99}
and to $1/4$ in an open cavity.\cite{Jalabert94,Blanter00,Oberholzer01,Oberholzer02}
For a recent review, see Ref.~\onlinecite{BlanterPhysRep00}.

Different to mesoscopic devices with normal electron reservoirs,
shot-noise can be enhanced in devices with superconducting leads
by virtue of the Andreev reflection process taking place at the
interface between a normal metal and
a superconductor.\cite{Averin96,Dieleman97,Hoss00,Nagaev01,Cron01,Roche01,Hoffmann02,Reulet02}
In some limiting cases, e.g. in the tunnelling and disordered limit,
the shot-noise can be doubled with respect to its normal
state value.\cite{Khlus87,deJong94,Muzykantskii94,Jehl00,Kozhevnikov00,Lefloch03}

In addition to measure shot-noise in a two-terminal geometry,
multi-terminal fluctuation measurements have been proposed.\cite{Buettiker92}
Whereas shot-noise corresponds to the autocorrelation of fluctuations,
cross-correlation measurements of fluctuations between different
leads provide a wealth of new experiments. As pointed out by
B{\"u}ttiker, exchange-correlations can for example be measured directly.\cite{Buettiker92}
In an attempt to go beyond conventional shot-noise measurements,
correlation measurements~\cite{Henny:HBT,Oliver99}
on electron beam-splitters~\cite{Liu98} were studied. The partitioning
of a `stream' of ferminons in a beam-splitter results in
negative correlations between the fluctuations measured on
the two output ports (anti-bunching). In contrast, bunching-like behavior
(positive correlation) has theoretically been predicted in multiterminal devices
in which at least one electrode is a superconductor.\cite{Anantram96,Torres99,Borlin02,Samuelsson02}
In the subgap region, charge is injected from the superconducting lead into the
device in correlated pairs of electrons, which - in the simplest picture -
may separate in the normal scattering region and exit at two different leads.
As a consequence, the current in the exit leads fluctuate in parallel.
However, it has been pointed out, that this picture is misleading, in particular
in the regime where the superconductor is strongly coupled to the normal region.
In this case, the normal region should rather be viewed as a proximity-induced
superconductor.\cite{Buettiker03} Positive correlations have
not been observed in mesoscopic devices until today.

Finally, we mention that the experimental quest for positive correlations is also
important for the new field of quantum computation and communication
in the solid state,~\cite{Loss98,DiVincenzo98}
in which entangled electrons play a crucial role.
A natural source of entanglement is found in superconductors in which
electrons are paired in a spin-singlet state. A source of entangled
electrons may therefore be based on a superconducting
injector.\cite{Choi00,Torres01,Recher01,Lesovik01,Recher02,Bena02,Chtchelkatchev02,Bouchiat03,Samuelsson02,Samuelsson02b}
Even more so, an electronic beam-splitter is capable of distinguishing
entangled electrons from single electrons.\cite{Burkard00,Loss00}

We have therefore focused our experimental research on the
fabrication of superconducting injectors (Nb) into a high-mobility
InAs-based two-dimensional electron gas (2DEG),\cite{Nitta92,Takayanagi95,Akazaki96,Neurohr96,Takagaki98,Toyoda99}
in which beam-splitters can be fabricated. In this article, we focus
on the fabrication of such devices in Sec.~\ref{exp}
and their characterization in terms of linear and
non-linear resistance in Sec.~\ref{R-measurements},
as well as shot-noise in Sec.~\ref{Noise-measurements}.
Near the superconductor-2DEG interface two
etched sub-micrometer constrictions define a
beam-splitter which divides the input current.
We study the shot-noise of the current from the superconductor to
one of the normal reservoirs and observe an enhancement
for bias currents corresponding to voltages below the
superconducting gap of the Nb contact. This enhancement is due to Andreev
reflections at the superconducting contact and disappears in
a magnetic field higher than the critical field of Nb.
Using the coherent scattering theory, we extensively
compare our measurements with different models in  Sec.~\ref{discussion}.
Our devices can best be described as composed of a highly transparent
S-N interface in series with a short scattering region, whose
size $L$ is comparable to the elastic mean-free path $l_e$.
It may therefore be named a quasi-ballistic S-N beam-splitter junction.

\section{\label{exp}Experiment}

The InAlAs/InGaAs he\-te\-ro\-struc\-ture was grown by molecular beam
epitaxy on a Fe-doped semi-insulating InP substrate. The 2DEG is
confined in a \mbox{$4$\,nm} wide InAs quantum well \mbox{$35$\,nm}
below the surface of the he\-te\-ro\-struc\-ture,
see Ref.~\onlinecite{Nitta92}  and Fig.~1a for details.
The substrate is first structured into  a \mbox{$50$\,$\mu$m} wide Hall bar
(MESA) by wet etching. Hall and Shubnikov-de Haas measurements (Fig.~1b)
then yield an electron density of \mbox{$n_e = 2.1 \cdot 10^{16}$\,m$^{-2}$}
and a mobility of \mbox{$\mu = 5.0$\,~m$^{2}$/Vs} for the 2DEG,
corresponding to a Fermi wavelength of \mbox{$\lambda_F= 18$\,nm} and
an elastic mean free path of \mbox{$l_e = 1.2$\,$\mu$m}.

The Nb electrode is defined by electron beam lithography at one side of
the MESA. First, the MESA is etched in the patterned electrode area to a depth
of \mbox{$\sim 50$\,~nm}. Then, the sample is mounted
in an evaporation chamber and rf-sputter cleaned. Without breaking the
vacuum, a \mbox{$80$\,nm} thick Nb film is subsequently deposited at an
angle of $30$ degrees to the horizontal. After lift-off, a \mbox{$50$\,$\mu$m}
wide superconductor-2DEG contact is obtained. A cross-section through
such a Nb contact is schematically shown in Fig.~1a.

\begin{figure}[htb]
Because our files are usually rejected (file size), we post it here without figures.
Please, refer to http://www.unibas.ch/phys-meso/Research/Papers/2004/Shot-Noise-Yjunction.pdf
\caption{\label{fig1} (a) Schematic side view of the Nb contact to
the InAs semiconductor he\-te\-ro\-struc\-ture and the corresponding
energy band diagram. (b) Longitudinal ($R_{xx}$) and Hall ($R_H$) resistance
measured on  this he\-te\-ro\-struc\-ture. (c) SEM picture of a sample (top view) with
a diagram of the measurement setup. The sample is current biased
through a series resistor and the voltage fluctuations are measured
with the aid of two sets of amplifiers whose outputs are cross-correlated.
$S$ denotes the power spectral density in units of V$^2$/Hz. }
\end{figure}

E-beam lithography is now used to reduce the macroscopic superconductor-2DEG contact
to sub-micron dimensions, see Fig.~1c.
This is achieved by etching trenches into the he\-te\-ro\-struc\-ture to a depth of \mbox{$60$\,nm}
below the surface, removing the conducting InAs quantum well.
Three trenches are etched, two vertical ones and one horizontal one,
which start at the nanometer-sized contact in front of the superconductor and
extend across the whole MESA. The vertical trenches have a width of about
\mbox{$\sim 100$\,nm} and are placed parallel and as close as possible
to the Nb interface at a distance of less than \mbox{$50$\,nm}.
The three-terminal junction consist thereafter of a \mbox{$300 \times 350$\,nm}
square area in the 2DEG which is bound on one side \mbox{($350$\,nm wide)} by the edge of the superconductor
and the other side by two constrictions leading to two macroscopic
normal electron reservoirs. The constrictions have a nominal width $w$ of \mbox{$170$\,nm},
corresponding to $N=2w/\lambda_F\approx 19$ conducting channels.
This part can be viewed as a beam-splitter for charge carriers
(Cooper pairs in the superconducting state), injected from the Nb contact.

The sample is mounted in a $^3$He cryostat with a base temperature of \mbox{$270$\,mK}.
Here, we restrict ourselves to two-terminal measurements as
schematically shown in Fig.~1c. The sample is current biased through a
\mbox{$1$\,M$\Omega$} series resistor thermally anchored at
the \mbox{$1$\,K} pot of the cryostat. The current is
determined by the DC bias voltage $U$, on which a small AC voltage is superimposed
in order to measure the differential resistance $dV/dI$.
All measurement lines are filtered at low temperature by
lossy microcoax cables and additional $\pi$-filters are used at
room temperature. Two ultra-low noise amplifiers (LI-75, NF
corporation) with a fixed gain of $100$, followed by two low-noise
amplifiers (Stanford SR560, operated at a nominal gain of $10$ or $100$),
are used to measure the voltage fluctuations
across the sample in parallel. All amplifiers are operated at room temperature
and powered by independent sets of batteries to minimize cross-talk.
The voltage signals from the amplifiers are then cross-correlated by a
spectrum analyzer (HP 89410A). This cross-correlation
technique~\cite{Glattli97} can eliminate (or greatly reduce)
the voltage noise contributions due to the two amplifiers, because
they ought to fluctuate in an uncorrelated manner.

\begin{figure}[htb]
\caption{\label{fig2} (a) Example of a measurement of the equilibrium (thermal) voltage
noise (spectral density $S_V$) versus temperature $T$
used to deduce the calibration parameters.
Here, the frequency and sample resistance were
\mbox{$f=110$\,kHz} and \mbox{$R\backsimeq 2$\,k$\Omega$}, respectively.
The thermal noise is linearly dependent on $T$ and the slope~\cite{explain1}
yields the attenuation factor $A$ of the signal. (b) The attenuation $A$ as a function
of $\omega R$ follows the dependence expected for a simple RC network, i.e.
$A(\omega)=(1+(\omega R C)^2)^{-1}$.}
\end{figure}

In order to measure shot-noise, which is a frequency independent
contribution, one has to ensure that $1/f$-noise can be neglected
at the highest bias currents. As a consequence, we have measured
the noise at rather large frequencies $f$ around \mbox{$50-200$\,kHz}.
In this window, $1/f$ noise can be neglected up to the highest
currents of \mbox{$\approx 2$\,$\mu$A}. Due to capacitances
in the whole circuit including the measurement lines, the
signal is damped.
The overall gain, including the frequency-dependent attenuation,
has to be carefully calibrated for each device separately.
This is done by measuring the equilibrium voltage noise
(i.e. the thermal noise), given by $S_{V} = 4 k_B T R$, as a function of
temperature $T$, as shown in Fig.~\ref{fig2}a.
Here, $k_B$ is the Boltzmann constant and $R$ is the linear-response
sample resistance (more precisely, the parallel connection of the sample resistance
with the series biasing resistor).
The measured voltage noise $S_{V_{1},{V_{2}}}$, including the amplifier noise, can be written as
(see Fig.~1c):

\begin{equation}
  \label{noise-calibration}
  S_{V_{1},{V_{2}}} = A(\omega) R^2 (S_I+S_{I,{\rm off}}) + S_{V,{\rm off}}
\end{equation}

Here, $S_I$ is the current noise of the sample, which in the calibration
procedure is of thermal origin only, i.e. $S_{I} = 4 k_B T / R$.
$S_{I,{\rm off}}$ denotes the current noise offset of
the two LI-75 amplifiers. This contribution cannot be eliminated by the
cross-correlation scheme. We also find a non-zero voltage noise offset $S_{V,{\rm off}}$
accounting for residual cross-talk between the amplifiers,
possibly due to spurious ground currents.
The nominal overall gain of either $10^3$ or $10^4$ of the amplifiers
has been divided off in the above Eq.~\ref{noise-calibration}. Hence, all
the quantities refer to `input' noise. Finally, $A(\omega=2\pi f)$ denotes the
frequency-dependent attenuation factor.

A typical calibration measurement at \mbox{$f=110$\,kHz} is shown in Fig.~2a.
The attenuation $A$ at this frequency is obtained from the slope
of $S_V(T)$ and the residual amplifier noise from the vertical offset of
the fitted linear dependence extrapolated to $T=0$.\cite{explain1}
The attenuation was measured
for different devices with varying resistances $R$, ranging between
$\approx 1.5$ and \mbox{$\approx 2.5$\,k$\Omega$} and frequencies
in the range of $50$ to \mbox{$\approx 200$\,kHz}.
$A(\omega)$ for a set of devices is shown in Fig.~2b to follow the expected damping for
a simple RC network, i.e. $A(\omega)=(1+(\omega RC)^2)^{-1}$.
The extracted capacitance of \mbox{$C=840$\,pF} is mainly due
to the filtering of the wires (microcoax-filters) and the
two input capacitances of the amplifiers.

The noise offset $S_{\rm off}$, extracted from the calibration procedure,
typically amounts to \mbox{$1\cdot 10^{-19}$\,V$^2$s}.
The current noise of a single LI-75 amplifier is specified to be \mbox{$ < 2 \cdot 10^{-28}$ \,A$^2$s}
and independently measured to be \mbox{$ < 8\cdot 10^{-28}$\,A$^2$s}, corresponding
to a voltage noise of \mbox{$< 3.2 \cdot 10 ^{-21}$\,V$^2$s} for a single amplifier
on a typical sample resistor of \mbox{$R=2$\,k$\Omega$}, or to
\mbox{$< 6.4 \cdot 10 ^{-21}$\,V$^2$s} for two amplifiers in parallel.
The offset current noise of the amplifiers is therefore at least an order of magnitude
smaller than the measured offset and can therefore not account for it.
Hence, the dominating part of the measured offset is
caused by residual voltage fluctuations and we set the amplifier current-offset to zero in
the following.
The voltage noise floor of a single LI-75 amplifiers is specified to
be \mbox{$1.4 \cdot 10 ^{-18}$\,V$^2$s} and independently measured (short circuit input)
to be \mbox{$2.5\cdot 10^{-18}$\,V$^2$s}, a value which is substantially
larger than the measured offset noise after the cross-correlation. The cross-correlation
technique therefore reduces the voltage fluctuations of the amplifiers
by as much as a factor of $25$.

The deduced calibration parameters are then used to extract the intrinsic
current shot-noise $S_I$ generated in a superconducting-2DEG junction from
the measured noise $S_{V_1,V_2}$ using Eq.~\ref{noise-calibration}.
It is important to emphasize, that $R$ in Eq.~\ref{noise-calibration} has to be replaced
by the differential resistance $dV/dI$ for the non-equilibrium measurement. This is crucial,
because of the non-linear current-voltage characteristic of these devices.

\section{\label{results}Results}

We measured the linear-response resistance $R$
as a function of temperature $T$, the differential resistance
$dV/dI$ and the spectral density of the voltage fluctuations
(the noise) as a function of bias current $I$, both at \mbox{$T=270$\,mK}.
We focus first on the resistance and then on the noise measurements.

\subsection{\label{R-measurements}Resistance measurements}

\begin{figure}[htb]
\caption{\label{fig3} Temperature dependent resistance
$R(T)$. Circles correspond to the measurements, whereas the
curves are calculated using the BTK model together with
a classical series resistor \mbox{$R_S=0$} (solid),
\mbox{$R_S=500$} (dashed), and \mbox{$R_S=1000$\,$\Omega$} (dotted).
$\Delta$ was fixed to \mbox{$1.14$\,meV} and the barrier
transparency $\Gamma$ was fitted, yielding \mbox{$\Gamma\approx 0.72$}.
The inset shows the superconducting transition of the Nb film
measured with two probes on a structured device. The transitions
of the Nb film and the sub-micron Nb contact are marked by arrows.}
\end{figure}

Fig.~3 shows the temperature dependence of the linear-response
resistance $R$ measured from the superconductor to one of the normal reservoirs,
as schematically shown in Fig.~1c. Above \mbox{$T=7.5$\,K}, the resistance
is constant, whereas it varies non-monotonically below. $R$ first drops
abruptly below \mbox{$7.5$\,K}, has a minimum at \mbox{$\approx 6$\,K}
and then starts to increase for lower temperatures. At the lowest
temperature, $R$ is \mbox{$\approx 8$\,\%} higher than $R(T>7.5\,{\rm K})$.
The drop at \mbox{$7.5$\,K} is identified with the
superconducting transition temperature $T_{c} $ of the junction.
The superconducting transition of the Nb film was also measured
via two contacts bonded to the Nb electrode of the actual device.
We found \mbox{$T_c=8.5$\,K} (inset of Fig.~3). A suppression of
$T_c$ in a film from its bulk value of \mbox{$9.5$\,K} is
commonly observed, as is a similar suppression of $T_c$
in micro-fabricated structures. The relative modest suppression
of \mbox{$\approx 1$\,K} is in agreement with previous work, see for
example Ref.~\onlinecite{Hoss00}.

The non-monotonic temperature dependence, which we observe in
Fig.~3, suggests that the superconductor-2DEG interface has an
intermediate transparency. This is qualitatively deduced by
referring to the BTK model of a superconductor-normal metal
junction.\cite{Blonder82} In this model scattering is exclusively
taking place at the junction interface described by a single parameter,
which is the transmission probability (transparency) of the
junction. This situation is referred to as the clean or ballistic
junction limit (also the BTK limit), as opposed to the case in
which additional scattering in the normal part of the junction is
introduced. If the junction has a low transparency
(tunnel junction), the resistance is expected to increase
exponentially fast at low temperature. On the other hand, if the
junction has a very high transparency, $R$ decreases monotonically
to reach half of its normal state value at the lowest
temperature. We neither see an exponential increase, nor a
monotonic decrease of $R$, suggesting intermediate transparency.

In the following, if we refer to the normal state resistance $R_N$, we
mean $R(\sim 8\,{\rm K})$, and if we refer to the resistance
in the superconducting state $R_S$, we mean $R(270\,m{\rm K})$.

The measured normal-state resistance $R_N$ of this device equals \mbox{$2.13$\,k$\Omega$}. It
is straightforward to compare the corresponding normal-state
conductance $G_N=R_N^{-1}$ with the Landauer formula,\cite{Imry_book} i.e.
with $G_N=(2e^2/h)N\Gamma$, where $N$ is the number of eigenchannels
with non-zero transmission eigenvalues ${\cal T}_n$ and $\Gamma$ the mean value
of ${\cal T}_n$. Taking $N$ to be $19$, as determined from the width
of the constrictions, yields $\Gamma \approx 0.32$ as
the average transmission coefficient of the entire device.
The resistance can have contributions from both the superconductor-2DEG
interface and the point contacts to the normal reservoirs. Therefore,
$\Gamma \approx 0.32$ must be seen as a lower bound for the S-N
interface transparency. This will be studied in greater detail
in section~\ref{discussion}.

In Fig.~3 are also shown calculated curves of $R(T)$. The solid
curve corresponds to the BTK model for a junction transparency of
\mbox{$72$\,\%}. The minimum of $R(T)$ is much more pronounced in the
calculated curve.
In an attempt to account for additional scattering, for example
at the constrictions of the beam-splitter, a classical series
resistor was added (dashed and dotted curves). This clearly improves
the overall matching, but strong deviations remain close to
$T_c$.

We mention that similar resistance values and temperature dependencies
were measured for several other samples.

\begin{figure}[htb]
\caption{\label{fig4} Voltage dependent differential resistance
$dV/dI(V)$ measured at \mbox{$T=270$\,mK}.
Circles correspond to the measurements, whereas the
curves are calculated using the BTK model together with
a classical series resistor \mbox{$R_S=0$} (solid),
\mbox{$R_S=500$} (dashed), and \mbox{$R_S=1000$\,$\Omega$}
(dotted). The parameters are similar to the ones
used in Fig.~2.
The full arrows point to the gap value $\Delta$ estimated from the transition
temperature $T_c$ using the standard BCS relation
\mbox{$\Delta=1.76 k_B T_c$}, whereas the open arrows point to
$\Delta=1.9 k_B T_c$, where the factor $1.9$ is known for bulk Nb.}
\end{figure}

We also measured the differential resistance $dV/dI$, which is shown
as a function of voltage $V$ in Fig.~4. What actually was measured is
$dV/dI$ as a function of bias current $I$. This data was converted to
the displayed voltage dependence by integration. Similar to the
temperature dependence, $dV/dI$ has a non-monotonic dependence. It first drops
for increasing voltage and shows a minimum (a dip) before
increasing again at higher voltages. The dip occurs close to the
gap value $\Delta$ of the superconductor. $\Delta$ is estimated
from the apparent transition temperature \mbox{$T_c=7.5$\,K} of the junction
using the zero-temperature BCS relation $\Delta = 1.76 k_BT_c$, yielding
\mbox{$\Delta = 1.14$\,meV} (black arrows). The agreement is even better
if we use instead of the BCS factor of $1.76$ for the ratio $\Delta/k_B T_c$
the factor $1.9$, which is the reported ratio for bulk Nb.
This yields \mbox{$\Delta = 1.23$\,meV} (open arrows).
Similar to $R(T)$, we used the BTK model to calculate the differential
resistance, which is shown as a solid curve. The dashed and dotted
curves correspond as before to the BTK model including a classical
resistor in series.
The theoretical curves display very pronounced
dips at $\pm\Delta$, which are apparently strongly damped in the
measurements. Unlike in the temperature dependent case, i.e.
$R(T)$, the series-resistor model improves the agreement only marginally. In
particular the strong dips are not removed.

\subsection{\label{Noise-measurements}Shot-noise measurements}

We measured the shot-noise from the superconductor to one
of the normal reservoirs of the sub-micrometer beam-splitter as schematically shown
in Fig.~1c. The measurement yields $S_{V_1 V_2}=S_V$ as a function
of bias current $I$. To obtain the intrinsic current noise $S_I(I)$ of the junction,
Eq.~\ref{noise-calibration} is applied using the calibration parameters as we
have described it in the experimental part of Sec.~\ref{exp}.
The result is shown in Fig.~5. It corresponds to the same sample, for which
$R(T)$ and $dV/dI(V)$ have been shown in Fig.~3 and Fig.~4, respectively.

\begin{figure}[htb]
\caption{\label{fig5} Power spectral density $S_I$ of the current noise
of a sub-micrometer S-N junction as a function of applied current $I$.
$S_I$ is extracted from the measured voltage noise $S_{V_{1},{V_{2}}}$ between the superconductor
and one of the normal reservoirs (see Fig.~1c) according to the Eq.~\ref{noise-calibration}.
A clear crossover from a large Fano factor $F_S$ at small bias currents to a reduced
Fano factor $F_N$ for large currents is observed. This crossover coincides with
gap $\Delta$ of the superconductor (open arrows).}
\end{figure}

The full temperature and voltage dependence of the power-spectral density $S_I$
of the current fluctuations can only be expressed in a simple analytical form
for a junction with a constant channel transmission coefficient ${\cal T}$. It is
given by~\cite{Buettiker92}
\begin{equation}
  S_I = \frac{4 (1-F)k_B T}{R} + F\cdot 2eI coth\left(\frac{eV}{2 k_BT}\right)
\end{equation}
where $F$ is known as the Fano factor and equals $1-{\cal T}$.

Noise measurements are generally analyzed in two limiting cases:
(a) for small applied voltages $eV << k_BT$, for which $S_I$ equals the
Johnson-Nyquist equilibrium noise (the thermal noise) $4k_BT/R$,
and (b) for large applied voltages $eV >> k_BT$, for which
a linear dependence of $S_I(I)$ is expected. In the limit of shot-noise, i.e. the latter
case, $S_I= F\cdot 2eI$ and it is the Fano factor $F$, which is the
central parameter that is deduced from such
measurements.\cite{Birk95,Reznikov95,Beenakker03}
$F=1$ for a junction in which all channels have low transmission eigenvalues, i.e. in
tunnel junctions.\cite{Birk95} In electronic devices in which charge is transported
by single electrons alone, the Fano factor can in general be written
as $F=\sum{{\cal }_n {\cal T}_n(1-{\cal T}_n)}/\sum{{\cal T}_n}$, which is always smaller or at most
equal to one. Hence, the \emph{suppression} of shot-noise in mesoscopic
devices has been a central focus of research during recent years. For a review we refer to
Ref.~\onlinecite{Beenakker03} and Ref.~\onlinecite{BlanterPhysRep00}. In contrast to `normal' conducting
devices, enhancement of shot-noise has recently been found in superconducting
devices, in S-I-S~\cite{Dieleman97} and S-N junctions,\cite{Jehl00,Lefloch03}
as well as in superconducting S-N-S links.\cite{Hoss00,Hoffmann02}
The two extreme cases of S-N junctions are the tunnel junction and the ballistic
junction. In the former, the noise in the
superconducting state is doubled \mbox{($F_S=2$)}
as compared to the normal state \mbox{($F_N=1$)}.\cite{Dieleman97,Lefloch03}
In the latter, shot-noise disappears completely, i.e. $F_S=F_N=0$.

The doubling of the shot-noise in the superconducting state may be interpreted as being caused by the
effective charge $e^{\star}$ of the
charge carriers,\cite{Khlus87,deJong94,Muzykantskii94,Averin96,Lefloch03}
which are Cooper pairs with $e^{\star}=2e$, provided the temperature and
the applied voltage are sufficiently small. One has to emphasize, that
the doubling of the shot-noise is not generic.\cite{Torres01} For a single channel
S-N junction with transparency ${\cal T}$, the ratio of the Fano factors
in the superconducting and normal state equals
$F_S/F_N = 8/(2-{\cal T})^2$, which - as mentioned before - can reach at most $2$.
If there are many channels with a distribution of
eigenvalues ${\cal T}_n$, the situation is different. For example, there
is a doubling from $F_N=1/3$ to $F_S=2/3$ in the diffusive case,\cite{Beenakker94,Jehl00,Kozhevnikov00}
but $F_N=1/4$~\cite{Jalabert94,Oberholzer01} increases to
$F_S=0.604$ in case of an open chaotic cavity with a superconducting and
normal terminal. The ratio in this case is even larger than $2$, i.e.
$F_S/F_N=2.4$.

The measured shot-noise in Fig.~5 clearly displays two regimes in which
$S_I (I)$ is nearly linear. In the low-current (low-voltage) regime,
the slope is larger than in the high-current (high-voltage) regime.
The crossover on the positive \mbox{($I>0$)} and negative  \mbox{($I<0$)} side
of the curve occurs at \mbox{$\approx 0.62$\,$\mu A$} and
\mbox{$\approx -0.78$\,$\mu A$}, corresponding to a voltage of
\mbox{$\approx 1.3$\,mV} and \mbox{$\approx -1.5$\,mV}, in reasonable agreement
with the value of the superconducting gap parameter \mbox{$\Delta/e=1.23$\,mV} (open arrows),
which we have deduced before. The agreement is good on the positive side, but
somewhat off on the negative side, where
the crossover appears to be shifted to a larger value. Asymmetries in the
crossover as well as in the Fano factors were seen in other samples too.
The low and high-bias slopes are identified with $F_S$ (low-currents) and $F_N$
(high currents). We deduce $F_S = 0.58 \pm 0.10$ and $F_N = 0.25 \pm 0.04$
(average of slopes for $I<0$ and $I>0$).
We note that the values of the Fano factors are considerably suppressed as compared to
the case of a weakly transparent S-N junction.

In contrast to conductivity measurements, from which
the average transmission probability can be deduced, measurements of
the shot-noise provide insight into the actual distribution
of the transmission eigenvalues, which helps to find the
correct description of the scattering problem of the actual device.
By making use of all measured parameters, the resistance in the normal and
superconducting state, as well as $F_S$ and $F_N$,
different models will be compared in detail in the last section.

\begin{figure}[htb]
\caption{\label{fig6} Fano factors versus perpendicular magnetic
field $B$ (symbols). The dashed curves are guides to the eyes.
As $B$ is increased, the enhancement of the Fano factor in
the superconducting relative to the normal state \mbox{(i.e. the ratio $F_S/F_N$)}
diminishes and fully disappears for \mbox{$B \agt 3$\,T}.
Note, that $F_N$ also slightly decreases as the field increases.
Inset:
The power spectral density $S_I$ of the current noise as a function of the bias current $I$
for \mbox{$B=0$}, $1$, $2$, and \mbox{$4$\,T}. The curves are shifted vertically
for clarity. The crossover (arrows) between the superconducting and normal state
shifts to lower voltages for increasing magnetic field as expected.}
\end{figure}

Finally, shot-noise measurements were also performed in
a perpendicular magnetic field $B$, see Fig.~6. It is seen that
the separation in two regimes, characterized by distinct Fano factors,
disappears around \mbox{$B=3$\,T}, corresponding to the critical field of the Nb contact,
which was measured independently.
Fig.~6 also shows that not only the Fano factor $F_S$ in the superconducting
state is suppressed, a decrease, though a smaller one, is also observed
in the normal state for $F_N$.
The origin is likely due to magnetic-field induced suppression
of backscattering in the semiconductor nanostructure,
a well known phenomenon in mesoscopic physics.\cite{Beenakker_VanHouten}
This observation proves that scattering is taking place
within the beam-splitter in zero magnetic field
adding up with the finite transparency of the
superconductor-2DEG interface to the whole scattering problem.
If we assume that ideally transmitting
edge states have formed at the highest field, the superconductor-2DEG interface
would have to account for the remaining Fano factor
of \mbox{$F_N = 0.16$} alone, yielding a transparency
of as much as \mbox{${\cal T}=1-F_N=0.84$} in a single channel model.
From reference transport measurements on Hall-bars we know that
the longitudinal resistance of the 2DEG displays pronounced
magnetic-field induced oscillations (Shubnikov-de Haas oscillations)
for \mbox{$B\alt 3$\,T}. Though the resistance minima do not yet reach zero,
clear quantum Hall plateaus are discernible, see Fig.~1b.
At \mbox{$4$\,T}, for example, the Hall measurements show that
$10$ Landau levels are occupied. Hence, the number of edge channels
is already smaller than the number of transporting channels
in zero magnetic field, which was estimated from the width
of the constrictions to be $19$. Since, transport follows the
edges in the quantum Hall regime, the \mbox{$84$\,\%} transmission
at the S-N interface must be seen as an upper bound for
the respective transmission probability in zero magnetic field.

In order to understand both the resistance and the shot-noise data,
we need to thoroughly compare our data with a model
consisting of a S-N contact with finite
transparency to which an additional scattering region is added.

\section{\label{discussion}Discussion and Modelling}

\begin{figure}[htb]
\caption{\label{scheme-models} Illustration of the two basic
models which we have used to analyze our data. (a) is based on a
wire and (b) on a cavity. The models are considered in different
regimes, but always in the limit of zero temperature. In (a) we
distinguish between the ballistic \mbox{($l_e\gg L$)},
intermediate \mbox{($l_e\sim L$)}, and diffusive \mbox{($l_e\ll
L$)} regime, whereas in (b) the cavity is assumed to be either
open on both sides (no barrier) or only open on one side with a
tunnelling barrier on the other side, described by its
transparency $\Gamma$.}
\end{figure}

We compare the data of one device with a set of models. These models are
schematically shown in Fig.~7.
Fig.~7a is the wire model with a fixed number of
channels $N$. Ideal contacts are assumed for the superconductor on
the left and normal metal on the right side. The barrier, which
may form in the processing of the Nb contact to the InAs 2DEG is
captured by a tunnelling barrier with transparency $\Gamma$. A
disordered region, parameterized by its length $L$ and elastic
scattering mean-free path $l_e$, can be included.
Fig.~7b is the cavity model. Here, both sides
can have different numbers of channels. This is in fact closer to
the real device geometry where the contact on the Nb side is wider
than the constrictions at the Y-branch. The contact on the right
is always assumed to be `open', meaning that its conductance is
equal to $NG_0$, where $G_0=2e^2/h$ is the quantum conductance,
whereas there may be a tunnelling barrier in the left contact in
order to model the effective transparency of the N-S contact. In
the following when we refer to `the normal state' we consider the
N-N case in which the superconductor is in the normal state.
Similarly, when we refer to `the superconducting state' we
consider the S-N case. Note, that in contrast to the real device
all the models have two terminals only. This simplification is
likely to introduce deviations, because the open third terminal
will add dephasing. However, neglecting dephasing (relaxation in
general) facilitates the comparison with theory greatly. Now, we
can use the machinery of mesoscopic physics to calculate the
conductances and shot-noise Fano factors in the normal and
superconducting state. It is one of the great hallmarks of
mesoscopic physics that these quantities can be calculated in the
coherent transport regime if the distribution $\rho({\cal T})$ of
transmission eigenvalues ${\cal T}$ for the particular device is known.

At zero temperature $T=0$ the respective equations for the
conductances $G_{(N,S)}$ and shot-noise powers $S_{(N,S)}$ in the
normal ($N$) and superconducting ($S$) state are:\cite{deJong97}
\begin{eqnarray}
\label{GN} G_N  & = & G_0 N \int_{0}^{1} d{\cal T} \rho({\cal T}){\cal T} \\
\label{GS} G_{S}  & = & G_0 N \int_{0}^{1} d{\cal T}\rho({\cal T})\frac{2{\cal T}^2}{(2-{\cal T})^2} \\
\label{SN} S_N &=&  S_0 N \int_{0}^{1} d{\cal T} \rho({\cal T}) {\cal T}(1-{\cal T}) \\
\label{SS} S_{S} &=& S_0 N \int_{0}^{1} d{\cal T}
\rho({\cal T})\frac{16{\cal T}^2(1-{\cal T})}{(2-{\cal T})^4}
\end{eqnarray}
where $S_0 = 2eV G_0$ and  $N$ is the total number of conducting
channels in the system. Even more so, general concepts have been
developed allowing to calculate the distribution function
$\rho({\cal T})$ for all models shown in Fig.~7.\cite{Belzig00}

The result of this comparison is summarized in table~\ref{t1}. In
the following we will go sequentially through the models and
discuss the assumptions and results. We focus on the quantities
$G_{(N,S)}$ and $F_{(N,S)}$ at zero
temperature. In case of the simplest models we will also compare
with the full temperature dependence of the conductance
$G(T)$ and the voltage dependence of the differential
conductance $dI/dV(V)$. The parameters $G_S$ and $G_N$ are deduced in
the experiment from the linear-response conductance measured at the smallest
temperature \mbox{$270$\,mK} and at \mbox{$\sim 8$\,K}, respectively.

The simplest possible model to compare with is a S-N junction in
which the normal part is ballistic. This problem was first
considered by Blonder, Tinkham and Klapwijk and is known as
the BTK model.\cite{Blonder82} In the BTK model of a S-N
interface, the junction is characterized by a single transmission
coefficient, i.e. $\rho({\cal T}) = \delta ({\cal T}-\Gamma)$. For $\Gamma=1$,
the junction resistance decreases with decreasing temperature and
the conductance is doubled at \mbox{$T =0$\,K} due to
Andreev reflection. In the opposite limit $\Gamma \ll 1$, Andreev
reflection is suppressed and the resistance increases
monotonically with decreasing temperature below $T_c$. The
comparison of the equations for $G_N$ and $G_S$ with the
experimental values yields $\Gamma =0.73$ and $N \approx 9$. We
can now use these two parameters to calculate the full temperature
and the non-linear voltage dependence of the conductance and compare
both with the measurements.
This is shown in Fig.~3 and 4 where the calculated curves are
the solid ones. Fig.~3 shows the temperature
dependence of the linear two-terminal resistance $R(T)$ and
Fig.~4 the differential resistance $dV/dI$ as a
function of voltage $V$, measured at \mbox{$T=270$\,mK}. As
imposed by this procedure the measured (circles) and calculated
(solid) curves in Fig.~3 match at zero temperature and at
(or above) $T_c$ in the normal state. Similarly, the measured (circles) and
calculated (dashed) curves in Fig.~4 match at zero
bias and approximately at the largest bias voltage of
\mbox{$|V|=4$\,mV} at which one closely approaches the normal
state. In the intermediate temperature and voltage regime
substantial deviations are found. The theory predicts a much
larger conductance increase in the intermediate regime than is
seen in the experiment. This is particularly striking in the
differential resistance where a strong dip (or a peak in the
conductance) is expected to occur near the superconducting gap
$\Delta$.

The experimentally observed strong damping of this conductance
peak near the superconducting gap
has also been seen in other work.\cite{Neurohr96,Jakob00}
It can be caused by pair-breaking due to inelastic scattering.
Even more so, the shape of the quasi-particle density-of-state in the
vicinity of \mbox{$\pm \Delta$}, which acquires singularities in the
BCS model, may strongly be damped at the interface between the
Nb and the 2DEG.\cite{Neurohr96}
The reason for the latter may be a disordered interface
caused by sputter cleaning or by partial oxidation. For
the former, we suspect that the second terminal of the Y-branch,
which has been left open, is a source of dephasing. Electrons at the Y-branch
can scatter into the drain contact, but may also be scattered into
the third terminal, from which they are reflected back but with
unknown phase. In addition, the large deviations in the
intermediate regime may also stem from the assumed model, which is likely
to be too simple. We will come back to this issue when we refine
the model. Let us now see whether the ballistic BTK model can
capture the shot-noise results, i.e. the measured Fano factors. In
case of an N-N barrier, the Fano factor is given by $F_N =
1-\Gamma$. The estimated $\Gamma = 0.73$ predicts $F_N = 0.27$,
which is consistent with the measured shot-noise Fano factor of
$0.25$. In the superconducting state, however, the theory for a
S-N barrier~\cite{deJong97} predicts $F_S = 8(1-\Gamma)/(2-\Gamma)^2
= 1.34$, whereas the measured Fano factor is substantially smaller
and amounts to $0.58$ only. We may also do the reverse and deduce
the transparency $\Gamma$  from the measured Fano factors instead.
$F_S = 0.58$ then implies $\Gamma = 0.91$
which is both inconsistent with the measured Fano factor in the
normal state $F_N$ and with the temperature dependence of the
resistance in Fig.~\ref{fig3}. Hence, the ballistic junction model
does not yield consistent values. This is not surprising, because
of the structured beam-splitter in front of the superconductor.
Each arm of the splitter is comprised of a relatively narrow
opening. Hence, parts of the eigenchannels emanating from the
Nb-2DEG interface must be back reflected at these exit ports. This
results in an additional voltage drop, i.e. in an additional
resistance. Since the sample is likely to be coherent this
resistance cannot simply be treated as a classical series
resistor. The whole structure composed of S-N interface, cavity
and exit leads need to be treated as one scattering problem. We
will discuss this latter on, but still try the classical series
resistor model as an additional test case next.
\begin{table*}
\caption{\label{t1}Comparison of the measured data, i.e. the
linear conductance $G_{(N,S)}$ and the shot-noise Fano factors
$F_{(N,S)}$ in the normal ($N$) and superconducting ($S$) state
with various models. Schematics for the models are shown in
Fig.~7a and b.}
\begin{ruledtabular}
  \begin{tabular}{cccccccc}
    & &\multicolumn{2}{c}{L $\ll l_e$} &\multicolumn{1}{c}{L $\gg l_e$} &\multicolumn{3}{c}{L $\sim l_e$} \\
    & \multicolumn{1}{c}{Measured} & \multicolumn{1}{c}{Ballistic} & \multicolumn{1}{c}{Ballistic} & \multicolumn{1}{c}{Diffusive} & \multicolumn{1}{c}{Open} & \multicolumn{1}{c}{Quasi-ballistic} & \multicolumn{1}{c}{Chaotic cavity} \\
    &          & \multicolumn{1}{c}{(BTK)} & \multicolumn{1}{c}{with $R_S$\footnotemark[1]}&        & \multicolumn{1}{c}{chaotic cavity} &  & \multicolumn{1}{c}{with Barrier} \\
    \colrule
    Fig.~7 & & a & a & a & b & a & b \\
    $G_N/G_0$       & $6.1 \pm 0.05$    & $6.1$     & $6.1$     & $5.8$     & $5.4$     & $6.1$     & $6.1$  \\
    $G_{S}/G_0$     & $5.5\pm 0.05$     & $5.5$     & $5.5$     & $5.8$     & $6.4$     & $5.5$     & $5.6$  \\
    $  F_N  $       & $0.25\pm 0.04$    & $0.27$    & $0.16$    & $0.33$    & $0.25$    & $0.36$    & $0.33$  \\
    $  F_S  $       & $0.58\pm 0.10$    & $1.34$    & $0.80$    & $0.67$    & $0.60$    & $0.77$    & $0.84$ \\
    $\Gamma$        &                   & $0.73$    & 0.72      & $1$       & $1$       & $0.55$    & $0.7$ \\
    $N$             &                   & $9$       & $11$      & $19$\footnotemark[2]  & $11$      & $17$      & $11$  \\
\end{tabular}
\end{ruledtabular}
\footnotetext[1]{series resistance, $R_S$ = 500$\Omega$}
\footnotetext[2]{the number of channels is fixed by the geometry, i.e. $N\sim 2w/\lambda_F$.}
\end{table*}

Fits to the measured two-terminal resistance $R(T)$ and
$dV/dI(V)$ including a classical resistor $R_S$ in series to the
S-N interface are shown in Fig.~3 and Fig.~4 for two values of $R_S$,
i.e. \mbox{$R_S=500$} and \mbox{$R_S=1000$\,$\Omega$}. It turns
out that if $R_S$ is increased, the fit of $R(T)$ improves in
the intermediate temperature regime. However, the width of the
zero-bias peak in $dV/dI$ broadens with increasing $R_S$, so that
the agreement gets worse here. A reasonable compromise is found
for \mbox{$R_S = 500$\,$\Omega$}. Using the conductance measurements
we deduce a junction transparency of $\Gamma = 0.72$ and obtain
for the number of channels $N \approx 11$ within this model. Because the
series resistor is a classical one it does not contribute to
non-equilibrium shot-noise. In order to deduce the Fano factor the
current fluctuations $S_I$ have to be plotted versus current $I$.
$S_I$ is obtained from the measured voltage fluctuations by
dividing $S_V$ with the total resistance $R=R_{SN}+R_S$ squared. In
the framework of this model this division is incorrect. Instead,
one should divide by $R_{SN}^2$, only. This now yields a
correction factor amounting to $(1+R_S/R_{SN})^2$, which has to be
applied to the measured data. For ease of comparison, we apply the
inverse $ 1/ \big(1+R_S/R_{SN} \big)^2 \approx 0.60$ to the model
calculation. As a result, the predicted Fano factor in the
superconducting state $F_S = 0.80$ is getting closer to the
measured value, but $F_N = 0.16$ is now clearly too small as
compared with the measured value. Adding a classical series
resistance improves somewhat the agreement between the experiment
and model of $R(T)$. It also relaxes slightly the large
discrepancy of the Fano factor in the superconducting state.
However, it is clear that this model is an oversimplification,
because the device is more than just one junction with a single
transparency and the whole device, including the cavity and beam-splitter
should be treated on equal footing.

A fixed transparency is a very idealized assumption, one which
never holds true in a practical multi-channel device. There are
many reasons why a distribution of transparencies has to be
considered: the junction interface is never perfectly homogeneous,
the sample has been structured and the boundaries may be rough on
the scale of the Fermi wavelength and there are dopants within the
he\-te\-ro\-struc\-ture. It is possible that the quality of the 2DEG was
degraded near the S-N interface during the sample processing, for
example, due to the Ar sputtering of the MESA prior to Nb
deposition.\cite{Neurohr96}
In addition, the narrow constrictions defining the
output ports must be seen as a scattering center. If we assume that
disorder is substantial, we are led to the diffusive regime, which
is another limiting case contrasting with the ballistic junction
limit discussed before. For a diffusive conductor, the
distribution of transmission eigenvalues $\rho({\cal T})$ is given by a
universal result $1/(2s {\cal T} \sqrt{1-{\cal T}})$,
where $s = L/l_e$.\cite{Beenakker92a} Using this distribution function yields
$G_{S}/G_N = 1$,\cite{Beenakker92b} $F_N=1/3$,\cite{Beenakker92a}
and $F_S=2/3$.\cite{deJong94} As can be seen
from the table, the agreement is much better, in particular for
the Fano factors, suggesting that elastic scattering must be
considered. However, the measured conductances are not equal in
the normal and superconducting state, i.e. $G_{S}/G_N = 1$, as
predicated by this model. Though the agreement is much better,
this model is an oversimplification too. We know that the
scattering-mean-free path in the bulk of the 2DEG is much larger
than the size of the nanostructure which is considered here.
In addition, the magnetic-field dependence of the Fano factor $F_N$ in the
normal state (Fig.~6) is inconsistent with a diffusive conductor.
One should therefore rather view the device as a cavity with three
terminals: a wide Nb one, and two narrow leads defined by the
constriction. This justifies to compare our data also to an open
chaotic cavity.

We only compare our data in table~\ref{t1} with the symmetric cavity,
because this is suggested by the measured Fano factor in the
normal state, which is found to be close to $F_N=0.25$. A
suppression factor of $1/4$ is the expected result for the
symmetric open cavity.\cite{Jalabert94,BlanterPhysRep00,Oberholzer01}
The distribution of transmission eigenvalues $\rho({\cal T})$ for a chaotic cavity, contacted
by two open leads each having $N$ ideally transmitting channels,
is given by another bimodal distribution function
$1/\pi\sqrt{{\cal T}(1-{\cal T})}$.\cite{Jalabert94,deJong97}
Using Eq.~\ref{GN}-\ref{SS} yields:
$G_N/G_0 = N/2$, $G_{S}/G_0 = ( 2 - \sqrt{2}) N$, $F_N=0.25$, and
$F_S=0.6036$. As can be seen from table~\ref{t1}, the
measured Fano factors compare very well with this model. On the
other hand, this model predicts $G_S > G_N$, whereas $G_S < G_N$ in
the experiment. We mention that $G_S > G_N$ also holds if the
cavity is allowed to be asymmetric. In fact, $G_S/G_N$ is minimal
for the symmetric cavity and reaches the well known factor of two
for strong asymmetries. This shows that we cannot cure the
deficiency in the conductances between theory and model just by
tuning the asymmetry alone. In an attempt to lower $G_S$ as
compared to $G_N$ we now further try to refine our model. There
are two refinements we can consider: We may start from the
`universial' diffusive case and ask the question what happens if
the elastic scattering mean-free path $l_e$ is increased up to the
point when $l_e$ becomes of the order of the device size (i.e. cavity
size). Secondly, we may add additional scattering by adding a
barrier to one side of the open cavity.

We first consider the `quasi-ballistic' case studied by De Jong
and Beenakker.\cite{deJong94,deJong97} In their model of a S-N
device, a tunnel barrier is inserted (which may be used to model
the quality of the contact itself) in series to a disordered
region of length $L$ in which the elastic scattering length is
$l_e$. De Jong and Beenakker were able to study the crossover from
the ballistic to the diffusive regime for an arbitrary ratio of $s = L/l_e$.
We have already considered the limiting cases $s=0$, which
is the ballistic BTK limit, and the universal diffusive case
$s\rightarrow\infty$. Interesting for us is the intermediate case
$s \sim 1$, which can be computed for both the normal and the
superconducting state using the scaling theory of the generalized
conductance.\cite{deJong94,Beenakker94} The numerical calculation
yields $\Gamma = 0.55$ and $N = 17$ for $s \sim 1$. De Jong and
Beenakker also showed that the shot-noise power can vary between
zero and twice the Poisson value, depending on the junction
parameters.\cite{deJong94} Using $\Gamma = 0.55$ and $N = 17$, we
obtain for the Fano factors $F_N = 0.36$ and $F_S = 0.77$.

In view of the real device geometry, a refinement of the open
cavity model is appealing too. The real device is asymmetric in
that the width of the contact at the Nb side is wider than the
constrictions at the exits. In addition, there is likely a barrier
at the interface of the 2DEG and the superconductor, the transparency of which has
been denoted by $\Gamma$ in the previous models. The simplest way to
calculate $\rho({\cal T})$ is to apply circuit theory \cite{Belzig00} to
the series connection of a tunnel junction with a quantum-point
contact (QPC). The tunnel junction is the element at the Nb side.
It is parameterized by its conductance $G_t$. The QPC models the
narrow constriction on the right side. It is parameterized by its
conductance $G=(2e^2/h)N$, i.e. by the number of (open) channels.
Though $\Gamma$ does not appear in the model explicitly (only the
ratio $G/G_t$ enters), it can be extracted from the fitted value
which we obtain for $G_t$. $G_t$ can be expressed as $G_0 \Gamma N
w_S/w_N$, where $w_{S,N}$ is the width of the 2DEG at the S and at
the N side, respectively. In trying to find the best match, we fix
the conductance in the normal state to the measured value and vary
$N$ to get the best agreement with all measured parameters. This
approach yields $N=11$, $\Gamma=0.7$, $F_N = 0.33$, and $F_S =
0.84$.

Let us summarize the results of all the models. One may say
that none yields perfect agreement in all four measured
parameters, i.e. $G_N$, $G_S$, $F_N$, and $F_S$. The most
realistic ones in terms of the actual geometry, i.e. the
quasi-ballistic and cavity with barrier models, yield reasonable
agreement in all parameters. The Fano factors are predicted to be
slightly larger than measured. In fact, this trend holds true for
all models considered. The measured Fano factors are
systematically smaller.
We suspect that the origin for this
discrepancy is found in the third terminal, i.e. the second
outgoing lead of the Y-branch, which was left open in the
measurements of the conductance and noise. Electrons entering into
this lead will relax and thermalize before being re-injected into
the device again. Relaxation in general reduces
shot-noise.\cite{Roukes85,Wellstood94,Nagaev95,Buettiker92,Henny97}
With regard to the number of channels the different
models predict $N=9\dots 17$ for the channel number in the
constriction. This is in fair agreement with an estimate of the
channel number based on the lithographic width and the Fermi
wavelength, yielding $N\sim 19$. It is quite reasonable that the
channel number deduced electrically turns out to be somewhat
smaller, because of depletion in the vicinity of the MESA after
etching.

\section{\label{con} Conclusions}

In summary, we have realized a mesoscopic superconductor-normal
beam-splitter geometry in a solid state hybrid system. We can
account for both the conductance and shot-noise data by modelling
the device as a highly transparent S-N interface connected in
series with a `short' scattering region, which is in the quasi-ballistic transport
regime. The scattering region is formed by the cavity in the 2DEG
between the S-N interface and the two constrictions forming the
electron beam-splitter. The shot-noise measured across
the superconductor and one arm of the beam-splitter
is enhanced relative to the normal state.
The respective Fano factors are in reasonable agreement with the
Landauer description (scattering problem) of coherent transport.
Residual deviations, in particular in the vicinity of the gap energy
in the differential conductance measurements, are likely due to
relaxation, a source of which is the second arm of the beam-splitter
which was left open in the reported experiments. Current fluctuations
can be suppressed by an extra terminal, even in the absence of a net (average) current.

Our devices are very well suited to explore positive
cross-correlations,\cite{Anantram96} as have recently been predicted in several
theoretical papers.\cite{Anantram96,Torres99,Lesovik01,Recher01,Borlin02,Samuelsson02}
Of these theoretical treatments, Ref.~\onlinecite{Samuelsson02} is in closest correspondence
with our experiments. In Ref.~\onlinecite{Samuelsson02}, an electron cavity
is connected to one superconducting and two normal leads via point contacts.
Positive correlations are predicted to appear for a dominant coupling to the superconducting lead.
The devices which we have studied in this work have roughly similar couplings to the S and N leads.
In the next step, one has to make use of the ability of semiconductors
to tune the transparency of the constrictions with additional electrodes (split gates),
which can be fabricated self-aligned with the etched trenches. This
would greatly help in the search for positive correlations in solid-state
nanostructures.


\begin{acknowledgments}
We thank G.~Burkard. M. Gr\"aber, P.~Recher, P.~Samuelsson, and C.~Strunk
for their fruitful discussions and acknowledge contributions to this
work by T.~Nussbaumer.
This work has been supported by the Swiss NFS, the NCCR on Nanoscience
and by the BBW (RTN `DIENOW' under the 5th framework EU programme).
\end{acknowledgments}


\clearpage


\begin{thebibliography}{99}
\bibitem{explain1}
The linear dependence of $S_V(T)-S_{\rm off}$ with temperature $T$, which we used in the
calibartion, is not exact, because $R$ is not constant, but temperature dependent as well.
$R$ varies in this temperature range by \mbox{$\pm \approx 10$\,\%}. For the calibration
an average value for $R$ was used.
\end{thebibliography}
\end{document}